\documentclass[a4paper,11pt]{article}
\usepackage{pos}
\usepackage{caption}
\usepackage{subcaption}
\usepackage{setspace}
\usepackage{caption}

\title{Benefits of Looking for Coincident Events, Taus, and Muons with the Askaryan Radio Array}
 \ShortTitle{Tau, Muon, and Multi-Station Events in ARA}

\author*[a]{Abigail Bishop}
\author[b]{Austin Cummings}
\author[b]{Ryan Krebs}
\author[c]{William Luszczak}

\affiliation[a]{Department of Physics, Wisconsin IceCube Astrophysics Center, University of Wisconsin-Madison,\\ Madison, WI 53706}
\affiliation[b]{Departments of Physics and Astronomy \& Astrophysics, Institute for Gravitation and the Cosmos, Pennsylvania State University,\\ University Park, PA 16802, USA}
\affiliation[c]{Department of Astronomy, Ohio State University,\\ Columbus, OH 43210, USA}

\onbehalf{for the ARA Collaboration \\{\normalsize \normalfont(a complete list of authors can be found at the end of the proceedings)}\\}

\emailAdd{abigail.bishop@wisc.edu}

\abstract{
  Ultra-High Energy (UHE) neutrinos over $10^{16}$ eV have yet to be observed but the Askaryan Radio Array (ARA) is one in-ice neutrino observatory attempting to make this discovery. 
  In anticipation of a thorough full-observatory and full-livetime neutrino search, we estimate how many neutrino events can be detected accounting for secondary interactions, which are typically ignored in UHE neutrino simulations. 
  Using the NuLeptonSim and PyREx simulation frameworks, we calculate the abundance and usefulness of cascades viewed by multiple ARA stations and observations made of taus, muons, and neutrinos generated during and after initial neutrino cascades. 
  Analyses that include these scenarios benefit from a considerable increase in effective area at key ARA neutrino energies, one example being a 30\% increase in ARA's effective area when simulating taus and muons produced in $10^{19}$ eV neutrino interactions. 
  These analysis techniques could be utilized by other in-ice radio neutrino observatories, as has been explored by NuRadioMC developers. 
  Our contribution showcases full simulation results of neutrinos with energies $3\times10^{17}$ - $10^{21}$ eV and visualizations of interesting triggered event topologies.
}

\ConferenceLogo{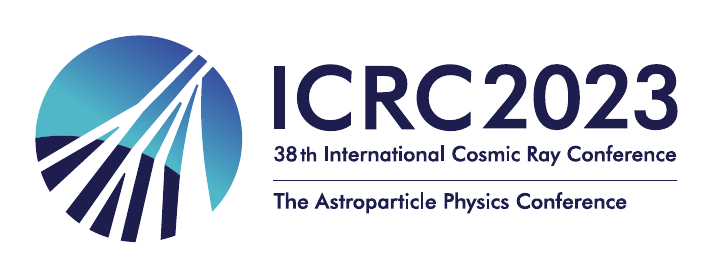}

\FullConference{
38th International Cosmic Ray Conference (ICRC2023)\\
  26 July - 3 August, 2023\\
  Nagoya, Japan}

\begin{document}
\maketitle

\section{Motivation}

  Ultra-high energy (UHE) neutrinos, those with energy $\geq 10^{17}$ eV, are believed to exist due to the observed flux of UHE cosmic rays \cite{crflux-auger, crflux-ta} and theoretical motivation of how these cosmic rays and other astrophysical sources can create UHE neutrinos \cite{crtheory1, crtheory2}. 
  Though a UHE neutrino has not been observed yet, there are models forecasting their flux \cite{koteramodel, vanvlietmodel} and many observatories attempting to observe them. 
  When a UHE neutrino interacts with a dielectric medium (like ice), Askaryan Radiation is emitted, forming a cone of coherent radio signal \cite{askaryan}.
  Conveniently, radio waves have favorable attenuation lengths in ice, allowing in-ice neutrino observatories to look for neutrinos over many cubic kilometers. 

  The Askaryan Radio Array (ARA) is one such neutrino observatory embedded in the glacier at the South Pole \cite{ara}.
  The array is composed of 5 stations, 2 kilometers apart, each with 8 horizontally and 8 vertically polarized radio antennas buried 100 or 200 meters in the ice. 
  The fifth station was fitted with 9 extra antennas installed in a dense, vertical line at the center of the station.
  Seven of these antennas are vertically polarized and form our Phased Array \cite{pa}.
  This upgrade has been shown to improve array sensitivity and has inspired subsequent in-ice UHE neutrino arrays like the Radio Neutrino Observatory of Greenland (RNO-G) \cite{rnog} and IceCube Gen2-Radio \cite{gen2}. 
  Although ARA has been taking data for over 10 years, only a quarter of our data has been analyzed \cite{ara2022}. 
  In these proceedings, ARA is presenting the framework for a full analysis covering data taken in all five stations from 2012 to 2022 with intentions to release full results within the near future \cite{5saICRC, A23ICRC}. 

  
  Neutrino interactions, charged current (CC) and neutral current (NC), result in an initial particle cascade that has been used to simulate the estimated sensitivities of individual stations. 
  In this work we focus on two additional detection capabilities that have not been considered in earlier analysis efforts. 
  The first is the use of interactions generated by outgoing leptons: tracks and decays from muons and taus from CC interactions and outgoing neutrinos from NC interactions.
  The probability that the outgoing neutrino will interact within the range of the arrays is quite small, but decays and tracks from muons and taus create cascades that could considerably increase ARA's effective area.
  The second detection capability is the observation of one event (including one or more cascade from one or more particles) with a secondary ARA station.  
  The ARA collaboration has calculated that 5\% of initial cascades from neutrinos with $10^{18}$ eV of energy trigger two or more stations \cite{aracoins}, but outgoing taus and muons could increase this.
  Additionally, if a cascade's signal is strong enough to meet analysis threshold in one station but too low on its own in a second station, the information gained from the triggered station could allow us to include sub-analysis threshold events in analysis. 
  Such scenarios are not explored in this exclusively trigger-level work; however, signals detected on additional stations could increase the overall event information available to us and may improve our sensitivity and event reconstruction.

  By only estimating our initial cascade sensitivity we may be underestimating the performance of our array and neglecting unique event topologies involving outgoing leptons, multiple triggered cascades from the same event, and multiple stations triggering overall.
  Having multiple energy depositions from the same event trigger or having multiple stations trigger on the same event could increase our ability to estimate the original neutrino's direction, energy, flavor, and initial cascade location. 
  This proceeding investigates the potential gains of looking for multi-station events in all five of our stations and the benefits of looking for taus and muons, in addition to primary neutrinos. 

\section{Simulations}

  We simulate initial neutrinos at discrete energies spaced logarithmically from $3\times10^{17}$ eV to $10^{21}$ eV entering Earth from all angles. 
  We simulate all subsequent interactions and cascades induced by outgoing leptons with energy greater than $10^{16}$ eV. 
  All resulting particle cascades that occur in a cylinder with a 15~km radius are passed into a signal simulation package where ARA's triggering is simulated.  

  The package that simulates the propagation of initial neutrinos, their interactions, and the propagations and interactions of all subsequent particles is NuLeptonSim \cite{nuleptonsim}.
  The model of Earth simulated follows the PREM model with a 2.8 kilometer layer of ice replacing the outermost 2.8 kilometers of Earth \cite{PREM}.
  Depending on their energy and flavor, between 1,000 and 10,000 primary neutrinos were injected along 100,000 unique trajectories that are evenly spread across Earth and that intersect with ARA's viewing region. 
  The distance traversed by each neutrino along its trajectory is randomly chosen according to the neutrino's cross section, which determines its interaction vertex. 
  All subsequent outgoing neutrinos are propagated the same way. 
  NuLeptonSim is similar to the existing PROPOSAL framework but has a faster implementation. 
  
  The outgoing muons and taus are propagated differently: particle cascades along muon and tau tracks are determined stochastically (in position and energy) along the charged lepton's trajectory until the particle randomly decays or has less than $10^{16}$ eV of energy remaining. 
  Outgoing neutrinos and charged leptons from muon and tau decays are also tracked, allowing us to incorporate $\nu_{\tau}\to\tau\to\nu_{\tau}$ regeneration in our study. 
  By simulating neutrinos all over Earth, we are also able to include particles that are created in interactions and decay far away but eventually travel into our trigger volume (traditionally, initial neutrino cascades are only simulated within the trigger volume). 
  All particle interactions and energy depositions in the 15-kilometer-radius trigger volume are characterized by location, energy, direction of travel, inelasticity, and interaction channel and passed to the next software. 

  Python for Radio Experiments, PyREx, simulates the radio signal emitted by the nearby cascades, then simulates the waveforms generated by ARA antennas upon receipt of those signals by all five of our stations\footnote{https://github.com/abigailbishop/pyrex}.
  PyREx does this by first ray tracing and attenuating the Askaryan signals from the position of an event to each radio antenna that can observe the cascade. 
  The electric field of the signal at each antenna is then convolved with the antenna's response function and modified through electronics filters until a voltage trace is achieved. 
  ARA's trigger is formed when the signal from multiple antennas exceeds a threshold determined by the power generated by a signal in comparison to the power from background noise.
  So, the pure signal voltage waveform and a pure noise voltage waveform are both convolved through a tunnel diode to get power traces for each. 
  For one station, if three or more antennas of the same polarization observe that the power generated from a cascade's signal is more than six times greater than the power generated by noise, the station triggers. 
  This is the classic ARA trigger; we do not consider Phased Array triggers in this study as this functionality is not built into PyREx yet.

\section{Results} 

  \subsection{Muons and Taus}

\begin{figure}
      \centering
      \includegraphics[width=4.25in]{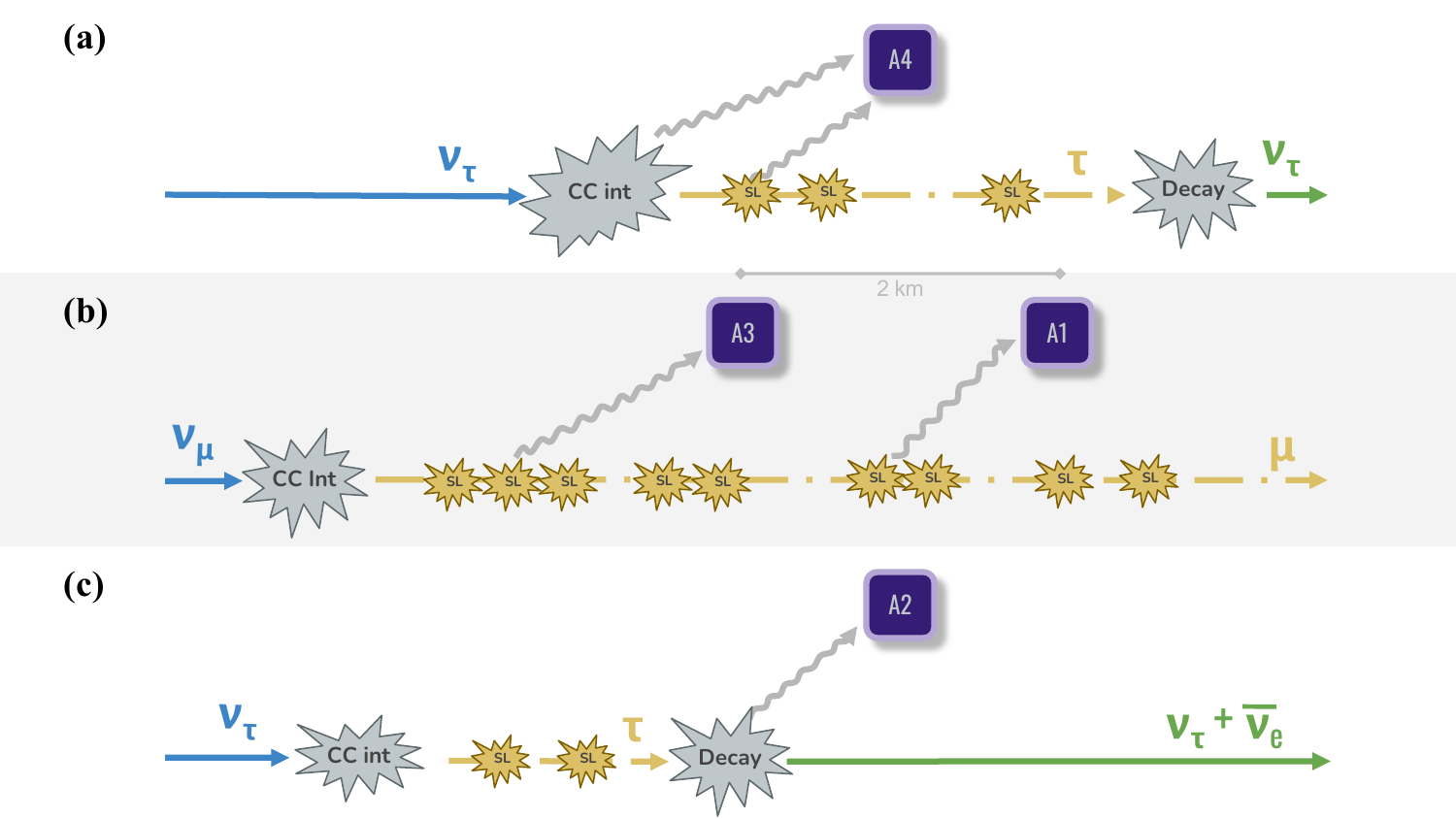}
      \caption{
        Cartoons of cascades from multiple particles in the same event triggering one station (a), 
        cascades from multiple stochastic losses (SL) from the same muon track triggering two different stations (b), 
        and a decay from a tau particle triggering one station (c). 
        Outgoing neutrinos from tau decays are depicted in green though they are unlikely to be observed.
      }
      \label{fig:eventtopos}
    \end{figure}

    Given ARA's multiple stations spread across multiple cubic kilometers of South Pole Ice, cascades from outgoing leptons are expected to have a considerable effect on our sensitivity.
    At ultra-high energies, taus have a decay length of tens-of-meters allowing us to observe tau tracks as well as tau decays \cite{nuradiomc2}. 
    This allows us to study a few interesting event topologies such as multi-particle events, muon and tau tracks, and muon and tau decays explained below and illustrated in Figure~\ref{fig:eventtopos}. 

    \begin{enumerate}

      \item Multi-cascade events are those where showers from one or more particles trigger our array.
      For example, a primary neutrino creates an initial cascade and the outgoing lepton generates secondary particle cascades in the ice that trigger the same ARA station. 
      Another would be multiple stochastic losses from the same muon triggering 2 ARA stations.
      (Figure~\ref{fig:eventtopos}a)

      \item For tracks, a muon or tau resulting from a charged current neutrino interaction 
      travels through our array's triggering region. 
      As the muon or tau travels, it stochastically sheds energy, initiating particle cascades that emit Askaryan Radiation which may be detected in more than one station.  
      Each deposit may be tens or hundreds of meters apart and each deposit, if observed, should have similar angular reconstruction and occur in time coincidence.
      This allows the observed energy depositions to be associated with each other. 
      (Figure~\ref{fig:eventtopos}b)

      \item The last interesting event topology is one in which a tau or muon decays and the associated particle cascades are observed by our array. 
      (Figure~\ref{fig:eventtopos}c)
      
    \end{enumerate}

    \begin{figure}
    
      \hspace{0.015\textwidth}
      \begin{subfigure}[t]{0.45\textwidth}
        \centering
        \includegraphics[width=2.5in]{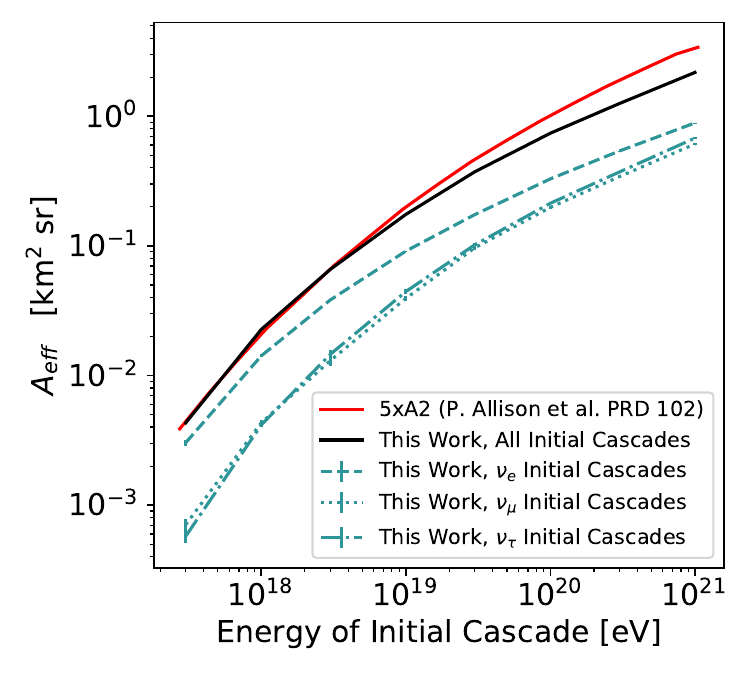}
        \caption{
          Effective Area of all events involving an initial neutrino cascade (in black) compared to previous ARA initial cascade results (in red) \cite{aracoins}. 
          The blue lines correspond to the effective area for different flavors of neutrinos. 
        }
        \label{fig:EAbyflavor}
      \end{subfigure}
      \hspace{0.035\textwidth}
      \begin{subfigure}[t]{0.45\textwidth}
        \centering
        \includegraphics[width=2.5in]{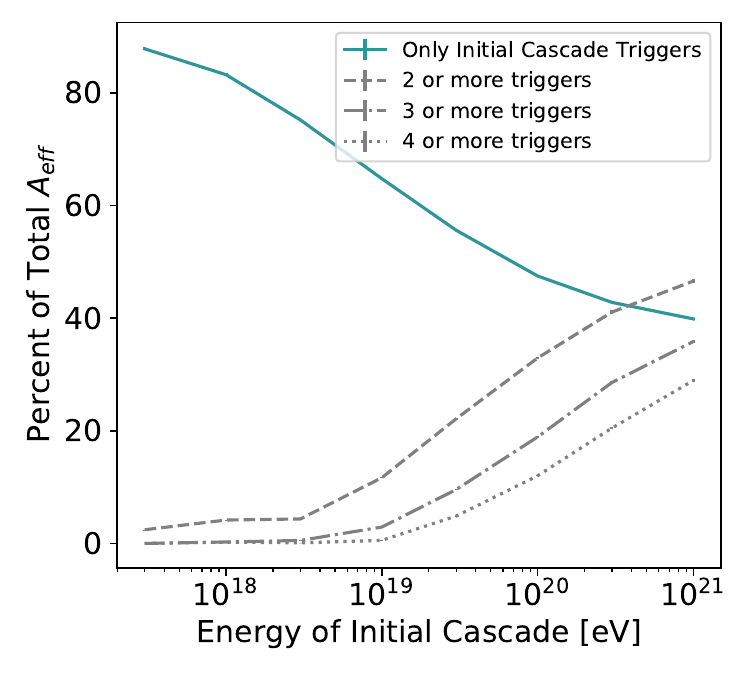}
        \caption{
          Percent of effective area composed by events where $only$ an initial cascade trigger (in blue) and events where two or more cascades from the same event trigger (gray lines) one or more ARA stations. 
        }
        \label{fig:EAbyNTriggers}
      \end{subfigure}
      
      \hspace{0.015\textwidth}
      \begin{subfigure}[t]{0.45\textwidth}
        \centering
        \includegraphics[width=2.5in]{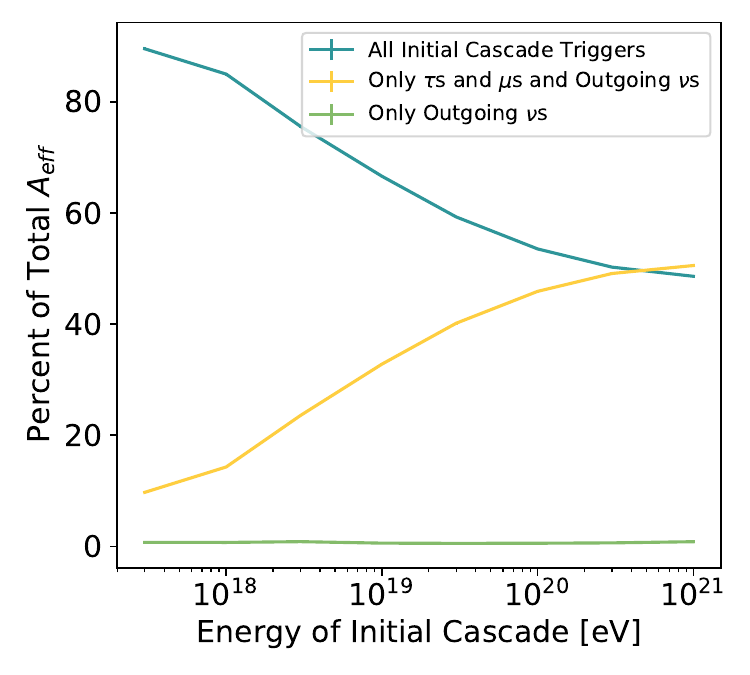}
        \caption{
          Percent of the Effective Area made up of: 
            all events including an initial cascade trigger (in blue, including events where cascades from an outgoing lepton triggered as well); 
            events where only cascades from taus, muons, and outgoing neutrinos trigger (in yellow);
            events where only outgoing neutrinos triggered (in green, all data points are between 0.5\% and 0.9\%). 
        }
        \label{fig:EAbyparticles}
      \end{subfigure}
      \hspace{0.035\textwidth}
      \begin{subfigure}[t]{0.45\textwidth}
        \centering
        \includegraphics[width=2.5in]{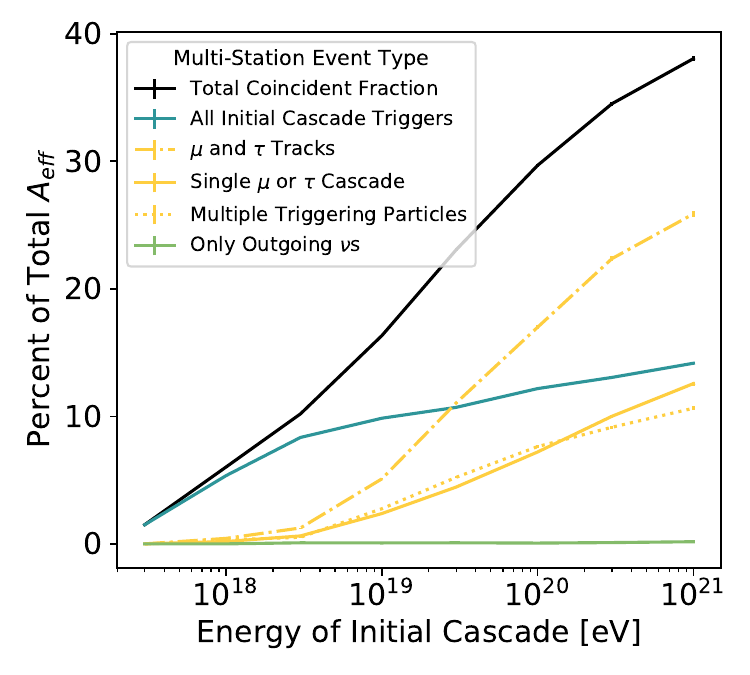}
        \caption{
          Percent of total ARA effective area from: 
            all multi-station events (in black), 
            initial neutrino cascades (in blue), 
            a single muon or tau cascade (from a track or decay) triggering multiple stations (in solid yellow), 
            multiple muon and tau cascades triggering different stations (in dash-dot yellow), 
            multiple particles from the same event triggering different stations (in dotted yellow), 
            and outgoing neutrinos (in green).
        }
        \label{fig:coinbreakdown-allflavor}
      \end{subfigure}

      \caption{
        ARA Effective Area with comparison to previous data and breakdowns by flavor, number of triggers per event, types of particles with triggering cascades, and types and abundance of events that trigger more than one station. The yellow curves in the bottom two plots reflect scenarios illustrated in Figure~\ref{fig:eventtopos}.
      }
      \label{fig:veffs}
    \end{figure}

    In the case of both tracks and decays, the tau or muon can originate inside or outside of our array's triggering region, so long as any cascades created exist within the triggering region. 
    Both the standard detector model, AraSim, and PyREx only simulate neutrinos originating within ARA's triggering region and estimate tau and muon decays at the site of the neutrino interaction \cite{arasim}.
    While the influence of tau and muon cascades on first generation arrays is small, making this a rational estimate, larger arrays that are currently under development should use a full simulation. 
    NuLeptonSim allows us to do so for this study by modeling secondary production for both muons and taus and propagation, using both continuous and stochastic energy losses. 
    Neutrino generation is performed in a larger volume, outgoing taus and muons are propagated through the ice using a fast algorithm that identifies tracks that pass through a smaller volume around the array. 
    Then the more time consuming signal generation can be performed for particles in a smaller volume, closer to the array~\cite{nuleptonsim}.
    NuRadioMC is a simulation framework that has a similar setting, using PROPOSAL for secondary production and propagation in a larger volume than considered for signal generation~\cite{nuradiomc1, nuradiomc2}. 

    The effective area results of our simulation are shown in Figure~\ref{fig:veffs}. 
    In Figure~\ref{fig:EAbyflavor}, we compare our signal-only initial cascade results to previous ARA signal+noise simulations of initial cascades and tau/muon approximations. 
    Although ARA's realistic use of noise and lepton approximations makes their effective areas larger than our calculations, our all-flavor initial cascade curve (red curve versus black curve) is reasonable. 
    Note that the ARA simulation we compare to is the effective area of 1 station multiplied by 5 (to match our 5-station simulation) which overestimates the effective area at high energies since this double counts multi-station observations \cite{araaeff}. 
    Considering we have verified our calculated effective areas, we now focus on the types and prevalence of events that make up the simulated effective area. 
    
    When we incorporate outgoing taus and muons, the effective area increases considerably. 
    As shown by the yellow curve in Figure~\ref{fig:EAbyparticles}, at $10^{19}$ eV, triggers on cascades from only tau tracks, muon tracks, and decays make up 30\% of ARA's effective area (20\% for those resulting from muon neutrino initial cascades, and 10\% from tau neutrinos).
    It's also clear from this plot that outgoing neutrinos from neutral current interactions or tau decays should not contribute greatly to our observations, as expected. 
    From the effective areas we calculated and the 2010 Kotera et al cosmogenic star formation rate flux (with $E_{max}=10^{21.5}$eV) \cite{koteramodel}, we expect up to 1.56 UHE neutrinos in our 25 station-year sample (0.13 via the 2019, 10\% proton, van Vliet et. al. flux \cite{vanvlietmodel}). 
    The strong performance of muon and tau cascades at high energies leads to a prediction of 0.52 (0.18) charged lepton events in our 25 station-years of data, or 1 event in 48 (139) station-years of data.
    This means cascades from charged leptons make up 25\% of all triggered events, which supports findings from the 2020 simulation study published by Garc{\'\i}a-Fern{\'a}ndez et al \cite{nuradiomcresult}. 
    
\begin{figure}
      \hspace{0.15\textwidth}
      \includegraphics[width=3.75in]{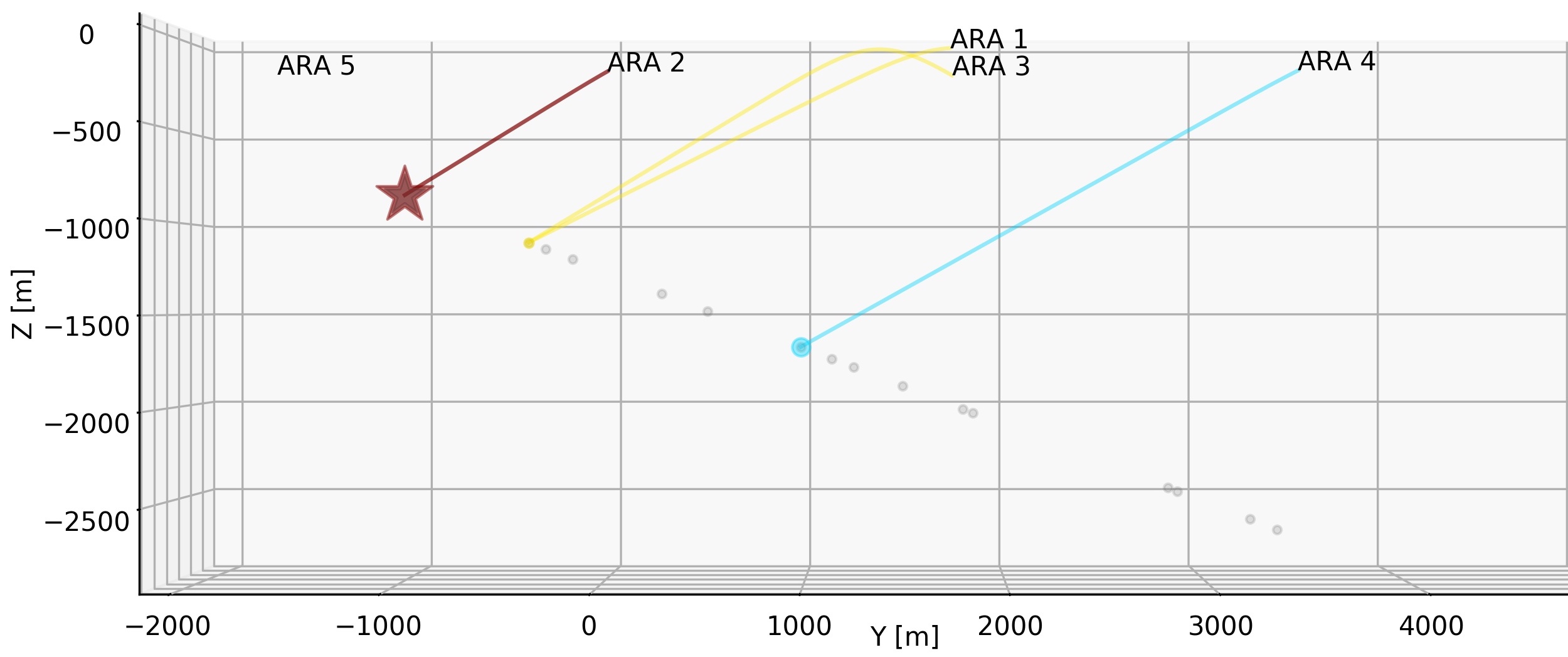}
      \caption{
        Example of an event where the initial tau neutrino charged current interaction (maroon star and ray) triggered ARA station 2. 
        Then the outgoing tau created a track of stochastic energy losses (grey dots) as the tau traveled downwards. 
        Two energy losses created cascades that triggered 3 more ARA stations (yellow and blue circles and rays).
      }
      \label{fig:eventdisplay}
    \end{figure}

    Figure~\ref{fig:EAbyNTriggers} shows that triggered events involving more than one triggering cascade (such as multiple energy depositions from tracks or an initial cascade and a subsequent decay) make up 13\% of triggered ARA events associated with $10^{19}$ eV initial cascades.
    An interesting event where three cascades from the same event triggered 4 out of 5 ARA stations is shown in Figure~\ref{fig:eventdisplay}. 
    This event involves a triggering charged current interaction from a $10^{19}$ eV primary tau neutrino followed by two triggering cascades from the outgoing tau's track of stochastic energy depositions, one of which triggered two stations at once.
    The predicted event observation count for events where two or more particles trigger is low: 0.17 events in 25 station-years of data according to the 2010 Kotera et al model (or 1 event in 147 station-years).
    However, for events originating from $10^{19}$ eV neutrino interactions, just 62\% involve only an initial cascade from a primary neutrino interaction. 
    The other 38\% of triggered events involve cascades from outgoing taus and muons showing the importance of their consideration in UHE neutrino observations, especially for larger in-ice radio arrays. 
    This may be an overestimate as our simulation does not account for DAQ downtime between events, meaning signal from a cascade may reach an ARA station when that station is busy recording a previous cascade and therefore not actively taking new data. 
    For triggers associated with $3\times10^{18}$ eV initial cascades the percent of triggered events including only an initial cascade is 73\% and is 85\% for $10^{18}$ eV. 
  
  \subsection{Multi-Station Events}
  
    Previous studies of events that trigger more than 1 ARA station concluded that, at $10^{18}$ eV, multi-station initial cascade triggers account for 5\% of events \cite{aracoins}. 
    Our multi-station results for initial cascades (Figure~\ref{fig:coinbreakdown-allflavor}, blue curve) are consistent with the previous measurement but the multi-station trigger prevalence increases to almost 7\% when considering tracks and decays from outgoing taus and muons.
    For events where the primary neutrino has $10^{19}$ eV of energy, the total multi-station contribution increases to 17\%, where almost half of these multi-station events occur due to the involvement of muon and/or tau cascades.  
    We predict there could be 0.25 multi-station events in 25 station-years of data according to the 2010 Kotera et al model.

    Looking for events that trigger more than one station, while interesting, is a challenge. 
    If we want to do fully integrated inter-station reconstruction, stations need to have synchronized clocks or that we know their relative timings. 
    We attempted to synchronize ARA station clocks using a White Rabbit networking switch at the IceCube Lab, but three of our stations had incompatible SFPs (electrical-to-optical converters). 
    
\section{Conclusion} 

  We estimated the impact of outgoing muons, outgoing taus, and multi-station events in an array that historically has focused on only initial cascades and single-station triggering events. 
  The trigger-level effective areas we calculated with our signal-only simulation predicts that the ARA 10-year dataset may contain one or two cosmogenic neutrinos. 
  While tracks and multi-station events could take decades to observe in our five-station array, a 35-station array like RNO-G or a hundreds-of-stations array like IceCube-Gen2-Radio will be able to accumulate more station-years of data in less time and could be able to see muons, taus, and multi-station events every few years once they are fully installed. 
  Not only do these arrays have more stations, but they also have next generation station hardware and designs that could allow for further increased likelihood for observing unique events. 

  \begin{center}
    \begin{minipage}{0.45\textwidth}
      \vspace{0.2in}
      \includegraphics[width=2.45in]{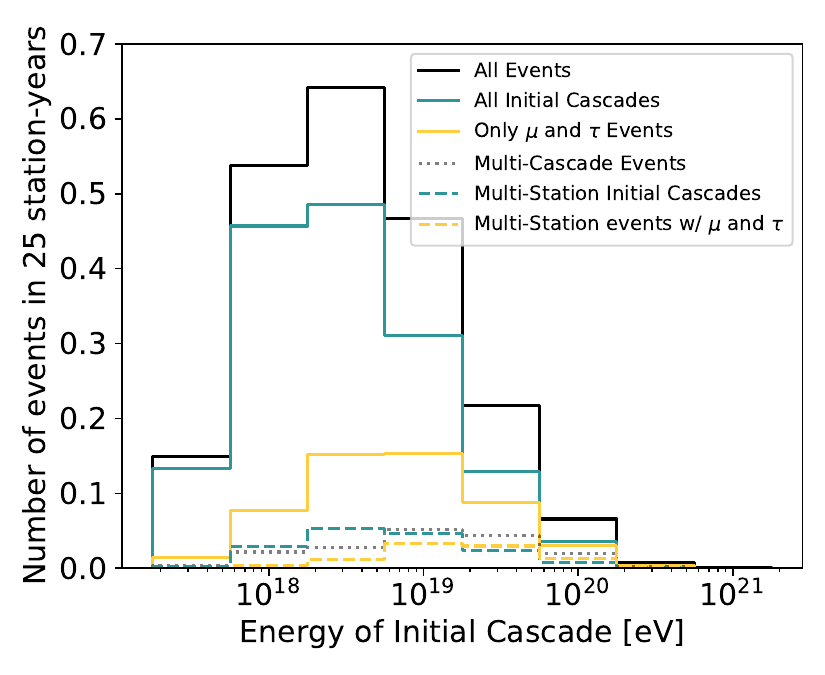}
      \label{fig:eventrates}
    \end{minipage}
    \begin{minipage}{0.45\textwidth}
      \footnotesize
      \begin{center}
        Number of events in 25 station-years:
        \begin{tabular}{| p{1.55in} | c |}
          \hline
          Event Type & $N_{\text{events}}$ \\ \hline \hline
          All Events & 2.09 \\ \hline
          All Initial Cascade Events & 1.56 \\ \hline
          Only $\mu$ and $\tau$ Events & 0.52 \\ \hline
          Multi-Cascade Events & 0.17 \\ \hline
          Multi-Station Initial Cascade Events & 0.16 \\ \hline
          Multi-Station events with $\mu$ and $\tau$ Cascades & 0.09 \\
          \hline
        \end{tabular}
      \end{center}
    \end{minipage}
    \captionof{figure}{Right: event rates calculated in centered, half decade bins from the effective areas for select event types and the 2010 Kotera et al model \cite{koteramodel}. Left: The total number of events integrated over all energies for each curve in the figure on the right.}
  \end{center}


\begingroup
\setstretch{0.8}
\setlength{\bibsep}{2.5pt}
\bibliographystyle{ICRC}
\bibliography{references}

\providecommand{\href}[2]{#2}\begingroup\raggedright\begin{thebibliography}{10}

\bibitem{crflux-auger}
{\bfseries Pierre Auger} Collaboration, P.~{Abreu} {\em et~al.}
  \href{http://dx.doi.org/10.1140/epjc/s10052-021-09700-w}{{\em EPJ C}
  {\bfseries 81} no.~11, (Nov., 2021) 966}.

\bibitem{crflux-ta}
{\bfseries Telescope Array} Collaboration, D.~{Ivanov} {\em et~al.} vol.~2019
  of {\em APS Meeting Abstracts}, p.~G08.002.
\newblock Jan., 2019.

\bibitem{crtheory1}
K.~Greisen \href{http://dx.doi.org/10.1103/PhysRevLett.16.748}{{\em Phys. Rev.
  Lett.} {\bfseries 16} (Apr, 1966) 748--750}.

\bibitem{crtheory2}
R.~{Engel}, D.~{Seckel}, and T.~{Stanev}
  \href{http://dx.doi.org/10.1103/PhysRevD.64.093010}{{\em PRD} {\bfseries 64}
  no.~9, (Nov., 2001) 093010}.

\bibitem{koteramodel}
K.~{Kotera}, D.~{Allard}, and A.~V. {Olinto}
  \href{http://dx.doi.org/10.1088/1475-7516/2010/10/013}{{\em JCAP} {\bfseries
  2010} no.~10, (Oct., 2010) 013}.

\bibitem{vanvlietmodel}
A.~{van Vliet}, R.~A. {Batista}, and J.~R. {H{\"o}randel}
  \href{http://dx.doi.org/10.1103/PhysRevD.100.021302}{{\em PRD} {\bfseries
  100} no.~2, (July, 2019) 021302}.

\bibitem{askaryan}
J.~{Alvarez-Mu{\~n}iz}, A.~{Romero-Wolf}, and E.~{Zas}
  \href{http://dx.doi.org/10.1103/PhysRevD.84.103003}{{\em PRD} {\bfseries 84}
  no.~10, (Nov., 2011) 103003}.

\bibitem{ara}
{\bfseries ARA} Collaboration, P.~{Allison} {\em et~al.}
  \href{http://dx.doi.org/10.1016/j.astropartphys.2011.11.010}{{\em
  Astroparticle Physics} {\bfseries 35} no.~7, (Feb., 2012) 457--477}.

\bibitem{pa}
{\bfseries ARA} Collaboration, P.~{Allison} {\em et~al.}
  \href{http://dx.doi.org/10.1103/PhysRevD.105.122006}{{\em PRD} {\bfseries
  105} no.~12, (June, 2022) 122006}.

\bibitem{rnog}
{\bfseries RNO-G} Collaboration, J.~A. {Aguilar} {\em et~al.}
  \href{http://dx.doi.org/10.1088/1748-0221/16/03/P03025}{{\em Jour. of Inst.}
  {\bfseries 16} no.~3, (Mar., 2021) P03025}.

\bibitem{gen2}
{\bfseries IceCube-Gen2} Collaboration, F.~G. {Schr{\"o}der}
  \href{http://dx.doi.org/10.48550/arXiv.2306.05900}{{\em arXiv e-prints}
  (June, 2023) arXiv:2306.05900}.

\bibitem{ara2022}
{\bfseries ARA} Collaboration, S.~Toscano {\em et~al.} {\em PoS} {\bfseries
  424} (05, 2022) 1.

\bibitem{5saICRC}
{\bfseries ARA} Collaboration, P.~Dasgupta, M.~Muzio, {\em et~al.} {\em PoS}
  {\bfseries 444} (8, 2023) 1226.

\bibitem{A23ICRC}
{\bfseries ARA} Collaboration, M.~Kim {\em et~al.} {\em PoS} {\bfseries 444}
  (8, 2023) 1148.

\bibitem{aracoins}
{\bfseries ARA} Collaboration, P.~Allison {\em et~al.}
  \href{http://dx.doi.org/10.1103/physrevd.93.082003}{{\em Physical Review D}
  {\bfseries 93} no.~8, (Apr, 2016) }.

\bibitem{nuleptonsim}
{\bfseries ARA} Collaboration, A.~Cummings {\em et~al.}
  \href{http://dx.doi.org/10.22323/1.424.0014}{{\em PoS} {\bfseries 424} (05,
  2022) 014}.

\bibitem{PREM}
A.~M. Dziewonski and D.~L. Anderson
  \href{http://dx.doi.org/https://doi.org/10.1016/0031-9201(81)90046-7}{{\em
  Physics of the Earth and Planetary Interiors} {\bfseries 25} no.~4, (1981)
  297--356}.

\bibitem{nuradiomc2}
C.~{Glaser} {\em et~al.}
  \href{http://dx.doi.org/10.1140/epjc/s10052-020-7612-8}{{\em European
  Physical Journal C} {\bfseries 80} no.~2, (Jan., 2020) 77}.

\bibitem{arasim}
{\bfseries ARA} Collaboration, P.~Allison {\em et~al.}
  \href{http://dx.doi.org/https://doi.org/10.1016/j.astropartphys.2015.04.006}{{\em
  Astroparticle Physics} {\bfseries 70} (2015) 62--80}.

\bibitem{nuradiomc1}
C.~{Glaser} {\em et~al.}
  \href{http://dx.doi.org/10.1140/epjc/s10052-019-6971-5}{{\em European
  Physical Journal C} {\bfseries 79} no.~6, (June, 2019) 464}.

\bibitem{araaeff}
{\bfseries ARA} Collaboration, P.~{Allison} {\em et~al.}
  \href{http://dx.doi.org/10.48550/arXiv.1912.00987}{{\em arXiv e-prints}
  (Dec., 2019) arXiv:1912.00987}.

\bibitem{nuradiomcresult}
D.~{Garc{\'\i}a-Fern{\'a}ndez}, A.~{Nelles}, and C.~{Glaser}
  \href{http://dx.doi.org/10.1103/PhysRevD.102.083011}{{\em PRD} {\bfseries
  102} no.~8, (Oct., 2020) 083011}.

\end{thebibliography}\endgroup
\endgroup

%

\clearpage

\section*{Full Author List: ARA Collaboration (July 18, 2023)}

\noindent
S.~Ali\textsuperscript{1},
P.~Allison\textsuperscript{2},
S.~Archambault\textsuperscript{3},
J.J.~Beatty\textsuperscript{2},
D.Z.~Besson\textsuperscript{1},
A.~Bishop\textsuperscript{4},
P.~Chen\textsuperscript{5},
Y.C.~Chen\textsuperscript{5},
B.A.~Clark\textsuperscript{6},
W.~Clay\textsuperscript{7},
A.~Connolly\textsuperscript{2},
K.~Couberly\textsuperscript{1},
L.~Cremonesi\textsuperscript{8},
A.~Cummings\textsuperscript{9}\textsuperscript{,}\textsuperscript{10}\textsuperscript{,}\textsuperscript{11},
P.~Dasgupta\textsuperscript{12},
R.~Debolt\textsuperscript{2},
S.~de~Kockere\textsuperscript{13},
K.D.~de~Vries\textsuperscript{13},
C.~Deaconu\textsuperscript{7},
M.~A.~DuVernois\textsuperscript{4},
J.~Flaherty\textsuperscript{2},
E.~Friedman\textsuperscript{6},
R.~Gaior\textsuperscript{3},
P.~Giri\textsuperscript{14},
J.~Hanson\textsuperscript{15},
N.~Harty\textsuperscript{16},
B.~Hendricks\textsuperscript{9}\textsuperscript{,}\textsuperscript{10},
K.D.~Hoffman\textsuperscript{6},
J.J.~Huang\textsuperscript{5},
M.-H.~Huang\textsuperscript{5}\textsuperscript{,}\textsuperscript{17},
K.~Hughes\textsuperscript{9}\textsuperscript{,}\textsuperscript{10}\textsuperscript{,}\textsuperscript{11},
A.~Ishihara\textsuperscript{3},
A.~Karle\textsuperscript{4},
J.L.~Kelley\textsuperscript{4},
K.-C.~Kim\textsuperscript{6},
M.-C.~Kim\textsuperscript{3},
I.~Kravchenko\textsuperscript{14},
R.~Krebs\textsuperscript{9}\textsuperscript{,}\textsuperscript{10},
C.Y.~Kuo\textsuperscript{5},
K.~Kurusu\textsuperscript{3},
U.A.~Latif\textsuperscript{13},
C.H.~Liu\textsuperscript{14},
T.C.~Liu\textsuperscript{5}\textsuperscript{,}\textsuperscript{18},
W.~Luszczak\textsuperscript{2},
K.~Mase\textsuperscript{3},
M.S.~Muzio\textsuperscript{9}\textsuperscript{,}\textsuperscript{10}\textsuperscript{,}\textsuperscript{11},
J.~Nam\textsuperscript{5},
R.J.~Nichol\textsuperscript{8},
A.~Novikov\textsuperscript{16},
A.~Nozdrina\textsuperscript{1},
E.~Oberla\textsuperscript{7},
Y.~Pan\textsuperscript{16},
C.~Pfendner\textsuperscript{19},
N.~Punsuebsay\textsuperscript{16},
J.~Roth\textsuperscript{16},
A.~Salcedo-Gomez\textsuperscript{2},
D.~Seckel\textsuperscript{16},
M.F.H.~Seikh\textsuperscript{1},
Y.-S.~Shiao\textsuperscript{5}\textsuperscript{,}\textsuperscript{20},
D.~Smith\textsuperscript{7},
S.~Toscano\textsuperscript{12},
J.~Torres\textsuperscript{2},
J.~Touart\textsuperscript{6},
N.~van~Eijndhoven\textsuperscript{13},
G.S.~Varner\textsuperscript{21},
A.~Vieregg\textsuperscript{7},
M.-Z.~Wang\textsuperscript{5},
S.-H.~Wang\textsuperscript{5},
S.A.~Wissel\textsuperscript{9}\textsuperscript{,}\textsuperscript{10}\textsuperscript{,}\textsuperscript{11},
C.~Xie\textsuperscript{8},
S.~Yoshida\textsuperscript{3},
R.~Young\textsuperscript{1}
\\
\\
\textsuperscript{1} Dept. of Physics and Astronomy, University of Kansas, Lawrence, KS 66045\\
\textsuperscript{2} Dept. of Physics, Center for Cosmology and AstroParticle Physics, The Ohio State University, Columbus, OH 43210\\
\textsuperscript{3} Dept. of Physics, Chiba University, Chiba, Japan\\
\textsuperscript{4} Dept. of Physics, University of Wisconsin-Madison, Madison,  WI 53706\\
\textsuperscript{5} Dept. of Physics, Grad. Inst. of Astrophys., Leung Center for Cosmology and Particle Astrophysics, National Taiwan University, Taipei, Taiwan\\
\textsuperscript{6} Dept. of Physics, University of Maryland, College Park, MD 20742\\
\textsuperscript{7} Dept. of Physics, Enrico Fermi Institue, Kavli Institute for Cosmological Physics, University of Chicago, Chicago, IL 60637\\
\textsuperscript{8} Dept. of Physics and Astronomy, University College London, London, United Kingdom\\
\textsuperscript{9} Center for Multi-Messenger Astrophysics, Institute for Gravitation and the Cosmos, Pennsylvania State University, University Park, PA 16802\\
\textsuperscript{10} Dept. of Physics, Pennsylvania State University, University Park, PA 16802\\
\textsuperscript{11} Dept. of Astronomy and Astrophysics, Pennsylvania State University, University Park, PA 16802\\
\textsuperscript{12} Universit\'{e} Libre de Bruxelles, Science Faculty CP230, B-1050 Brussels, Belgium\\
\textsuperscript{13} Vrije Universiteit Brussel, Brussels, Belgium\\
\textsuperscript{14} Dept. of Physics and Astronomy, University of Nebraska, Lincoln, Nebraska 68588\\
\textsuperscript{15} Dept. Physics and Astronomy, Whittier College, Whittier, CA 90602\\
\textsuperscript{16} Dept. of Physics, University of Delaware, Newark, DE 19716\\
\textsuperscript{17} Dept. of Energy Engineering, National United University, Miaoli, Taiwan\\
\textsuperscript{18} Dept. of Applied Physics, National Pingtung University, Pingtung City, Pingtung County 900393, Taiwan\\
\textsuperscript{19} Dept. of Physics and Astronomy, Denison University, Granville, Ohio 43023\\
\textsuperscript{20} National Nano Device Laboratories, Hsinchu 300, Taiwan\\
\textsuperscript{21} Dept. of Physics and Astronomy, University of Hawaii, Manoa, HI 96822\\

\section*{Acknowledgements}

\noindent
The ARA Collaboration is grateful to support from the National Science Foundation through Award 2013134.
The ARA Collaboration
designed, constructed, and now operates the ARA detectors. We would like to thank IceCube and specifically the winterovers for the support in operating the
detector. Data processing and calibration, Monte Carlo
simulations of the detector and of theoretical models
and data analyses were performed by a large number
of collaboration members, who also discussed and approved the scientific results presented here. We are
thankful to the Raytheon Polar Services Corporation,
Lockheed Martin, and the Antarctic Support Contractor
for field support and enabling our work on the harshest continent. We are thankful to the National Science Foundation (NSF) Office of Polar Programs and
Physics Division for funding support. We further thank
the Taiwan National Science Councils Vanguard Program NSC 92-2628-M-002-09 and the Belgian F.R.S.-
FNRS Grant 4.4508.01 and FWO. 
K. Hughes thanks the NSF for
support through the Graduate Research Fellowship Program Award DGE-1746045. B. A. Clark thanks the NSF
for support through the Astronomy and Astrophysics
Postdoctoral Fellowship under Award 1903885, as well
as the Institute for Cyber-Enabled Research at Michigan State University. A. Connolly thanks the NSF for
Award 1806923 and 2209588, and also acknowledges the Ohio Supercomputer Center. S. A. Wissel thanks the NSF for support through CAREER Award 2033500.
A. Vieregg thanks the Sloan Foundation and the Research Corporation for Science Advancement, the Research Computing Center and the Kavli Institute for Cosmological Physics at the University of Chicago for the resources they provided. R. Nichol thanks the Leverhulme
Trust for their support. K.D. de Vries is supported by
European Research Council under the European Unions
Horizon research and innovation program (grant agreement 763 No 805486). D. Besson, I. Kravchenko, and D. Seckel thank the NSF for support through the IceCube EPSCoR Initiative (Award ID 2019597). M.S. Muzio thanks the NSF for support through the MPS-ASCEND Postdoctoral Fellowship under Award 2138121.

\end{document}